# A TEM study of Si-SiO$_2$ interfaces in silicon nanodevices

Paul Spizzirri [1], Sergey Rubinov [2], Eric Gauja [3], Laurens Willems van Beveren [3]
Rolf Brenner [3] and Steven Prawer [1]

[1]*Centre of Excellence for Quantum Computer Technology, School of Physics, University of Melbourne, Melbourne, Victoria, 3010.*
[2]*The Bio21 Molecular Science and Biotechnology Institute, University of Melbourne, Melbourne, Victoria, 3010.*
[3]*Centre of Excellence for Quantum Computer Technology, School of Physics, University of New South Wales, Sydney, New South Wales, 2052.*

**Abstract**

The fabrication of micro- and nano-scale silicon electronic devices requires precision lithography and controlled processing to ensure that the electronic properties of the device are optimized. Importantly, the Si–SiO$_2$ interface plays a crucial role in defining these properties. While transmission electron microscopy (TEM) can be used to observe the device architecture, substrate / contact crystallinity and interfacial roughness, the preparation and isolation of the device active area is problematic. In this work, we describe the use of focussed ion beam technologies to isolate and *trench-cut* targeted device structures for subsequent TEM analysis. Architectures studied include radio frequency, single electron transistors and electrically detected, magnetic resonance devices that have also undergone ion implantation, rapid thermal and forming gas anneals.

**Introduction**

Current trends in semiconductor device fabrication have seen the realisation of sub-100nm lithographic processing in silicon. While many of the processing steps involved are now considered routine, there are implications for smaller device geometries and their interactions with neighboring atoms/interfaces. This is most evident in systems where the active elements comprise only a few atoms and where interfaces like that of Si-SiO$_2$, play a significant role in the operation of the device (e.g. MOS).

Work is being undertaken in this laboratory to develop a solid state quantum computer (QC) which is particularly sensitive to its electronic environment including that of the device substrate. Using shallow donor placement strategies, phosphorus ions are located within tens of nanometers from the substrate surface which is functionalized with MOS circuitry. The resulting number of processing steps that such a device undergoes before measurement can be significant and the underlying structures fabricated are very small (i.e. < 1 micron).

The importance of having analytical techniques capable of reporting on the structure/function of nano-electronic devices cannot be overstated, however finding those that will allow the fabrication pathway to remain intact for study is problematic. While state-of-the-art electron microscopic tools are well suited to the task of device analysis, it has been the development of focused ion beam (FIB) technologies that has solved the problem of sample isolation/preparation for analysis. In this work, we describe the use of electron and focused ion beam techniques for the identification and isolation of solid state structures for high resolution TEM analysis. Areas of interest for this analysis include: measurement of the gate oxide thickness after processing, observation of interfacial roughness and defects or damage to the silicon caused by processing.

**Experimental Details**

Devices were fabricated using conventional photo- and electron-beam lithographic techniques to create MOS structures on silicon. Oxide layers (field and device) were thermally grown in a triple wall quartz furnace and aluminium contacts were electron beam deposited. A ~100nm gold protective layer was sputter deposited onto the sample prior to ion beam processing to protect underlying structures. Electron beam imaging, which allowed identification of the area of interest and focused ion beam milling were both performed on an FEI NOVA dual beam FIB/SEM. The gallium LMIS of the FIB was also used to pattern a protective platinum layer onto the area being prepared for *liftout*. This is necessary to protect surface structures (i.e. sample integrity) from the gallium beam which can mill away structures as well as being implanted resulting in amorphised near surface regions. Once the membrane was removed, samples were positioned atop a carbon membrane and imaged using an FEI TECNAI TF20 TEM.

**Sample Preparation**

One example of a device that underwent FIB preparation and TEM analysis is the radio frequency, single electron transistor (RF-SET) shown in Figure 1. This device has a number of aluminium contacts patterned onto a high quality (device) Si-SiO$_2$ surface. Two red dots are depicted to indicate the location of the two phosphorus donors which were implanted. The area circled (blue) encompasses a ground plane which is the area of interest for the ensuing TEM analysis.

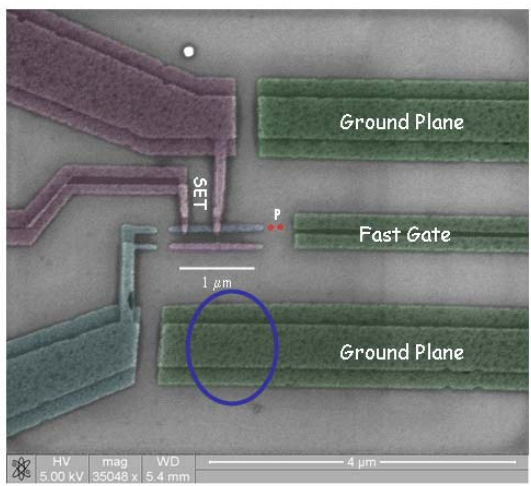

Figure 1. False coloured SEM image of the RF-SET device imaged prior to liftout. The area to be analysed is circled (blue).

Shown in Figure 2 is a lower magnification SEM image of the same device following Au coating and Pt deposition which defines the liftout region. The field oxide and device oxide areas are also depicted for clarity along with the photolithographic wirebond contacts (i.e. gold coloured). Figure 3 shows the finished membrane ready for removal and analysis using the TEM.

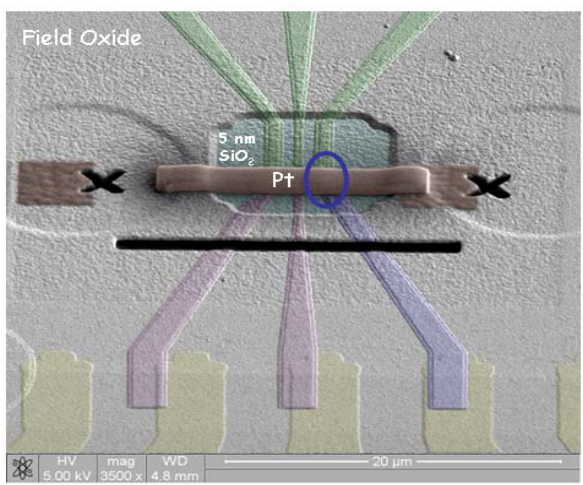

Figure 2. False coloured SEM image of the gold coated RF-SET with the protective Pt coated area imaged prior to membrane milling. The area of interest is circled (blue).

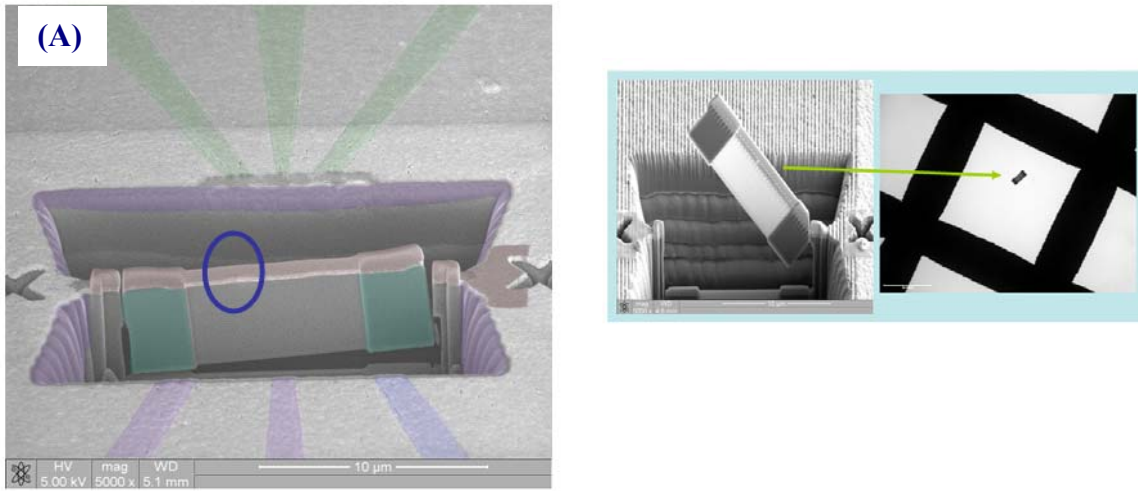

Figure 3. False coloured SEM images of (A) the liftout region (membrane) following FIB milling and (B) a membrane undergoing transfer for TEM analysis. The analysis area of interest is circled in (A) (blue).

**Interface Analysis Using TEM**

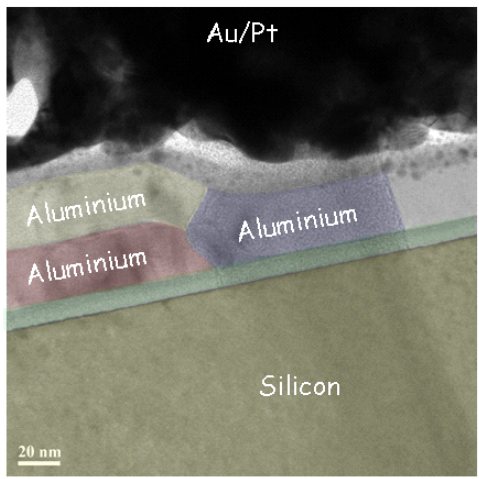

Figure 3. False coloured, medium resolution TEM image of the Si-SiO$_2$ interface with aluminium contacts.

Interface analysis was performed to characterize the: SiO$_2$ layer, Si-SiO$_2$ interface, aluminium schottky contacts and underlying silicon for dislocations or damage arising from processing. Shown in Figure 3 is a cross sectional view of the Si – SiO$_2$ – Al$_2$O$_3$ - Al interfaces. The aluminium grains (crystal), which have an interfacial oxide, are evident in the image as is the insulating SiO$_2$ layer. The Au and Pt overlayers provide contrast in the regions above the device surface.

Depicted in Figure 4 is a higher resolution image of the same region where it is clear that all of the interfaces are sharp and well defined suggesting that there is good process control. Secondly, the high quality device oxide, which is supposed to be ~5nm thick, is actually ~8.5nm thick. This difference in thickness is significant and has implications for the successful implantation and activation of dopant ions when using low ion energies. Finally, it is worth noting that there are no defects evident in the underlying silicon.

The quality of the Si-SiO$_2$ interface was also examined and compared to areas where field oxides (~200nm thick) had previously been grown and etched back with a high quality thermal oxide (5nm) subsequently regrown. This is standard practice in low noise MOS device fabrication, used to reduce leakage currents. Shown in Figure 5 is one area of a high quality SiO$_2$ layer which is ~5nm thick and exhibits a sharp interface. In comparison, the image shown in Figure 6 shows a less sharp Si-SiO$_2$ interface which has also undergone field oxide etchback and device oxide regrowth. For this sample, there is some evidence of interfacial roughening (< 1nm). This type of interface may result in a reduced performance for MOS devices so it is important to understand the mechanisms by which these interfaces are formed.

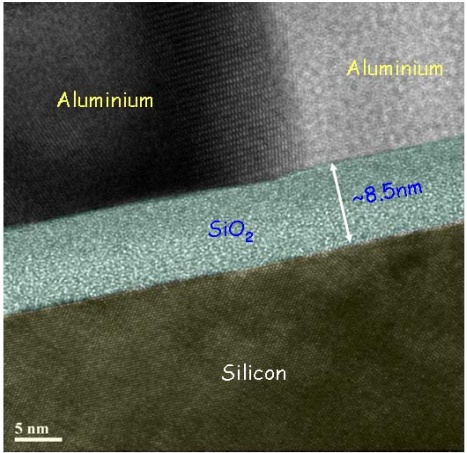

Figure 4. False coloured, high resolution TEM image of the Si-SiO$_2$-Al interfaces.

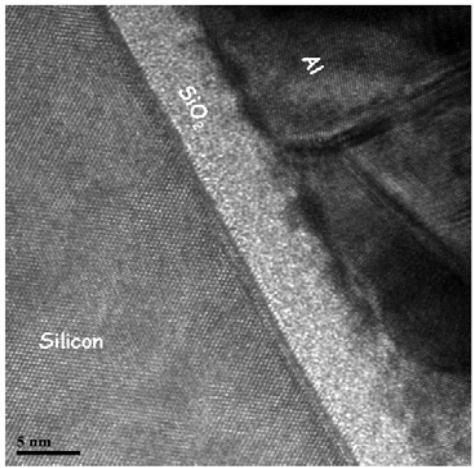

Figure 5. High resolution TEM image of an Si-SiO$_2$ interface with aluminium contacts.

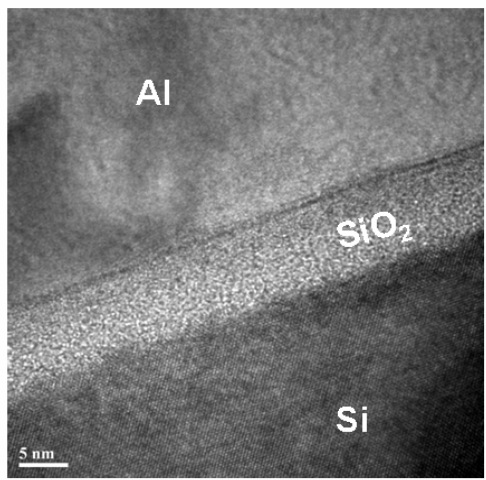

Figure 6. High resolution TEM image of a Si-SiO$_2$-Al interface that has undergone field oxide etchback and thermal oxide regrowth with some evidence of interfacial roughening.

## Conclusions

**F**IB prepared TEM samples were used in this work to study the resulting architecture of nano-scale structures used for quantum measurement. They were dissected with sub-micron precision from larger areas and issues surrounding the final silicon dioxide thickness were easily identified. Some interfacial roughening was also observed and may be associated with field oxide etchback and thermal oxide regrowth processes. This effect is still being investigated. These techniques ultimately confirmed the final device architecture and provided information which can be used to feedback into the fabrication process to optimize device geometries.

## Acknowledgements

The work is funded by the Australian Research Council, the Australian Government and US Army Research Office under Contract No. W911NF-04-1-0290.